
\input phyzzx
\tolerance=10000

\def\cref#1{\rlap,\attach{#1}}
\def\pref#1{\rlap.\attach{#1}}

\def\ie{{\it i.e.}}

\def\e{{\rm e}}

\def\half{{1\over 2}}

\def\pint{{1 \over 2\pi}\int d^2x\,}
\def\sqg{\sqrt{-g}}
\def\ephi{\e^{-2\phi}}
\def\eph{\e^{-\phi}}

\def\eprho{\e^{2\rho}}

\def\inv#1{\lbrack #1 \rbrack_{\rm inv.}}
\def\zebar{\bar\zeta}
\def\psbar{\bar\psi}

{\hsize=17.5truecm \leftskip=9.5cm
{NDA-FP-8/92, OCHA--PP-30}\par
\vskip -4mm
{September 1992}\par}
\title{Dilatonic Supergravity in Two Dimensions and
the Disappearance of Quantum Black Hole}
\author{Shin'ichi Nojiri}
\address{Department of Mathematics and Physics}
\address{National Defense Academy}
\address{Hashirimizu, Yokosuka 239, JAPAN}
\author{Ichiro Oda}
\address{Faculty of Science, Department of Physics}
\address{Ochanomizu University}
\address{1-1 Otsuka 2, Bunkyo-ku, Tokyo 112, JAPAN}

\abstract{We analyze a supergravity theory coupled to a dilaton
and superconformal matters in two dimensions.
This theory is classically soluble and we find
all the solutions appeared in Callan, Giddings, Harvey and Strominger's
dilatonic gravity also satisfy the constraints and the equations of motion in
this supersymmetric theory.
We quantize this theory by following the procedure of Distler, Hlousek
and Kawai.
In the quantum action, the cosmological term is renormalized to vanish.
As a result, any solution corresponding to classical black hole does not
appear in the quantum theory, which should be compared with the
non-supersymmetric case.}

\endpage

\REF\iii{C.G. Callan, S.B. Giddings, J.A. Harvey
and A. Strominger\journal Phys.Rev. &D45 (92) R1005}
\REF\i{S.W. Hawking\journal Comm.Math.Phys. &43 (75) 199}
\REF\vi{J.G. Russo, L. Susskind and L. Thorlacius,
preprint SU-ITP-92-4 (1992)}
\REF\vii{L. Susskind and L. Thorlacius, preprint SU-ITP-92-12 (1992)}
\REF\axi{S.W. Hawking\journal Phys.Rev.Lett. &69 (92) 638}
\REF\xii{T. Banks, A. Dabholkar, M.R. Douglas
and M. O'Loughlin\journal Phys.Rev. &D45 (92) 3607}
\REF\xviii{A. Bilal and C.G. Callan, preprint PUPT-1320 (1992)}
\REF\xix{S.P. de Alwis\journal Phys.Lett. &B289 (92) 278}
\REF\xx{K. Hamada, preprint UT-Komaba 92-7}
\REF\xvii{J.G. Russo, L. Susskind and L. Thorlacius,
preprint SU-ITP-92-17 (1992)}
\REF\xxvii{S.W. Hawking and S.A. Stewart, Univ. of Cambridge preprint (1992) }
\REF\xxxii{S.P. de Alwis, preprint COLO-HEP-288 (1992) }
\REF\yi{S.B. Giddings and A. Strominger, preprint UCSBTH-92-28(1992)}
\REF\xxviii{J.G. Russo and L. Susskind and L. Thorlacius, preprint UTTG-19-92,
SU-ITP-92-24 (1992)}
\REF\xxxxv{O. Lechtenfeld and C. Nappi\journal Phys.Lett. &B288 (92) 72}
\REF\xxxxvi{E. Elizalde and S.D. Odintsov, preprint UB-ECM-PF 92/ (1992)}
\REF\xvi{S. Nojiri and I. Oda, preprint NDA-FP-5/92, OCHA-PP-26
(1992), to be published in {\sl Phys.Lett.}{\bf B}, NDA-FP-6/92, OCHA-PP-27
(1992)}
\REF\xxv{A.M. Polyakov\journal Phys.Lett. &103B (81) 207}
\REF\xxx{F. David\journal Mod.Phys.Lett. &A3 (88) 1651}
\REF\xxix{J. Distler and H. Kawai\journal Nucl.Phys. &B321 (89) 509}
\REF\xxxi{J. Distler, Z. Hlousek and H. Kawai\journal Int.J of Mod.Phys. &A5
(90) 391}
\REF\xvii{K. Higashijima, T. Uematsu and Y.Z. Yu\journal Phys.Lett.&139B
(84) 161}
\REF\xxi{T. Uematsu\journal Z.Phys. &C29 (85) 143}
\REF\xxii{M. Hayashi, S. Nojiri and S. Uehara\journal Z.Phys. &C31 (86) 561}
\REF\xxiii{T. Uematsu\journal Z.Phys. &C32 (86) 33}
\REF\xiv{S. Nojiri\journal Phys.Lett.&B274 (92) 41}
\REF\xxvi{A.M. Polyakov\journal Phys.Lett. &103B (81) 291}
\REF\xxiv{M.M. Nojiri and S. Nojiri\journal Prog.Theor.Phys. &76 (86) 733}
\REF\xxxx{S. Nojiri\journal Prog.Theor.Phys. &77 (87) 159\journal
Phys.Rev. &D35 (87) 2466}




\chapter{Introduction}

The dilaton gravity theory proposed by Callan, Giddings, Harvey and
Strominger\refmark\iii (CGHS) is very instructive for the understanding
of black hole physics.
Especially, the problems associated with Hawking radiation\refmark\i
have been discussed by using this toy model\pref{\vi - \xvi}
In the original paper by CGHS, the quantum effects, such as the Hawking
radiation and its back reaction of the metric, were expected to be
described by adding correction term, which only comes from the conformal
anomaly\cref\xxv to the classical action.
Several authors\cref{\xviii - \xx} however, have claimed that the procedure
of David\refmark\xxx and of Distler and Kawai\refmark\xxix is necessary when
we quantize this theory consistently.
By using this procedure, it has been found\refmark{\xxxii , \yi} that the
quantum theory has no lower bound in energy and it has been conjectured
that this problem will be resolved by supersymmetrizing the theory.
In this paper, we propose a supergravity theory coupled to a dilaton in two
dimensions.
This theory is also classically soluble and we find all the solutions found
in Callan, Giddings, Harvey and Strominger's dilatonic gravity also satisfy
the constraints and the equations of motion in this supersymmetric theory.
We quantize this theory by following the procedure of Distler, Hlousek
and Kawai\pref\xxxi
In the quantum action, the cosmological term is renormalized to
vanish.
As a result, any solution corresponding to classical black hole does not
appear in the quantum theory.
This might tell that supersymmetry would forbid the existence of black hole
in quantum theory even in higher dimensions.
In the next section, we propose the classical action of dilatonic supergravity
by using the tensor calculus by Higashijima, Uematsu, Yu\cref{\xvii, \xxi}
which is based on conformal supergravity\pref{\xxi - \xxiii}
We show that all the solutions, including black hole solutions, in CGHS
theory are also solution of this supersymmetric theory.
Here it is interesting to observe that the cosmological constant is always
positive semi-definite as in CGHS model. This situation is analogous to that
of four dimensional supergravity, which can be constructed only in the
anti-de Sitter space.
In Section 3, we quantize this action following the procedure of Distler,
Hlousek and Kawai. The cosmological term cannot appear if we require that
the quantum action has superconformal symmetry. This tells that, in quantum
theory, there is not any solution corresponding to black hole solutions in
the classical theory. By bosonizing the fermion fields, we find that the
equations of motion which are obtained from the effective action of the
quantum theory are Liouville equations.
The last section is devoted to summary and discussion.

\chapter{Classical Black Hole}

We start from the following action.
This action will describe the effective action of superstring in two
dimensional black hole background\pref\xiv
We use the notations and the tensor calculus in the papers by Higashijima,
Uematsu, Yu\pref{\xvii, \xxi}\foot{The definition of the scalar curvature
$R$ in Refs.\xvii\ and \xxi\ is different from that in Ref.\iii\ by sign;
$R=-R^{\rm CGHS}$.}

$$\eqalign{
S=&\pint \Bigl(-2\inv{\tilde\Phi^2\otimes W}
-4\inv{\tilde\Phi\otimes T(\tilde\Phi)}
+4\lambda\inv{\tilde\Phi^2}\cr
&+\sum_i^N\half\inv{\tilde\Sigma_i\otimes T(\tilde\Sigma_i)} \Bigr) \cr
=&\pint e\Bigl\{-R\varphi^2+2S(\varphi f'+\zebar\zeta)
-4\varphi\zebar\sigma^{\mu\nu}\psi_{\mu\nu} \cr
&+4g^{\mu\nu}\partial_\mu\varphi\partial_\nu\varphi
+4\zebar\gamma^\mu\partial_\mu\zeta-4{f'}^2
-4\psbar_\nu\gamma^\mu\gamma^\nu\zeta\partial_\mu A
-\half\zebar\zeta\psbar_\nu\gamma^\mu\gamma^\nu\psi_\mu \cr
&+4\lambda\Bigl(2\varphi f'-\zebar\zeta+\varphi\psbar_\mu\gamma^\mu\zeta
+\half\varphi^2\psbar_\mu\sigma^{\mu\nu}\psi_\nu+S\varphi^2\Bigr)\cr
&+\sum_i^N\Bigl(-\half g^{\mu\nu}\partial_\mu a_i\partial_\nu a_i
-\half\bar\xi_i\gamma^\mu\partial_\mu\xi_i+\half{G'}_i^2 \cr
&+\half\psbar_\nu\gamma^\mu\gamma^\nu\xi_i\partial_\mu a_i
-{1 \over 16}\bar\xi_i\xi_i\psbar_\nu\gamma^\mu\gamma^\nu\psi_\mu\Bigr)\Bigr\}
}\eqn\ei$$
Here $\Sigma_i$'s are matter scalar multiplets
$\Sigma_i=(a_i, \xi_i, {G'}_i)$, $i=1,\cdots,N$.
A scalar multiplet $\tilde\Phi=(\varphi , \zeta, f')$ is given in terms of
a dilaton multiplet $\Phi=(\phi,\chi,F')$ by
$$\tilde\Phi\equiv\e^{-\Phi}
=(\eph,-\eph\chi,-\eph(F+\half\bar\chi\chi))\ . \eqn\eii$$
$W$ is a curvature multiplet,
$$W=(S,\eta,-S^2+\half R-\half\psbar^\mu\gamma^\nu\psi_{\mu\nu}
+{1 \over 4}S\psbar^\mu\psi_\mu)\ .\eqn\eiia$$
Here $\eta$ and $\psi_{\mu\nu}$ are defined by,
$$\eqalign{\eta\equiv &-\half S\gamma^\mu\psi_\mu
+{1 \over 2}e^{-1}\epsilon^{\mu\nu}\gamma_5\psi_{\mu\nu}\ , \cr
\psi_{\mu\nu}\equiv & D_\mu\psi_\nu-D_\nu\psi_\mu\ .}\eqn\eiib$$
$T(\Sigma)$ is a kinetic multiplet which is defined for a scalar multiplet
$\Sigma=(a, \xi, G')$ and
$\inv{\Sigma}$ expresses the invariant Lagrangian density which is
given by
$$\inv{\Sigma}\equiv e\lbrack
G'+\half\psbar_\mu\gamma^\mu\xi
+\half a\psbar_\mu\sigma^{\mu\nu}\psi_\nu+Sa \rbrack
\ . \eqn\ea$$
The action \ei\ is, by construction, invariant under the following local
supersymmetry transformation,
$$\eqalign{
\delta e_\mu^a=& \bar\epsilon\gamma^a\psi_\mu \cr
\delta \psi_\mu=&2(\partial_\mu+\half\omega_\mu\gamma_5+\half\gamma_\mu S)
\epsilon \cr
\delta S=&-\half S\bar\epsilon\gamma^\mu\psi_\mu
+\half i e^{-1}\epsilon^{\mu\nu}\bar\epsilon\gamma_5\psi_{\mu\nu} \cr
\delta\phi=&\bar\epsilon\chi \cr
\delta\chi=&\Bigl\{F'+\gamma^\mu(\partial_\mu\phi
-\half\psbar_\mu\chi)\Bigr\}\epsilon \cr
\delta F'=&\bar\epsilon\gamma^\mu\Bigl\{\Bigl(\partial_\mu
+\half\omega_\mu\gamma_5\Bigr)\chi \cr
&-\half\gamma^\nu\Bigl(\partial_\nu A
-\half\psbar_\nu\chi\Bigr)\psi_\mu-\half F'\psi_\mu\Bigr\} \cr
\delta a_i=&\bar\epsilon\xi_i \cr
\delta\xi_i=&\Bigl\{{G'}_i+\gamma^\mu(\partial_\mu a_i
-\half\psbar_\mu\xi_i)\Bigr\}\epsilon
\cr
\delta {G'}_i=&\bar\epsilon\gamma^\mu\Bigl\{\Bigl(\partial_\mu
+\half\omega_\mu\gamma_5\Bigr)\xi_i \cr
&-\half\gamma^\nu\Bigl(\partial_\nu \phi
-\half\psbar_\nu\xi_i\Bigr)\psi_\mu-\half {G'}_i\psi_\mu\Bigr\}\ .}\eqn\eei$$
Here $\epsilon$ is an anti-commuting spinor parameter of local supersymmetry
transformation and $\omega_\mu$ is the spin connection and given by
$$\omega_\mu=-ie^{-1}e_{a\mu}\epsilon^{\lambda\nu}\partial_\lambda e^a_\nu
-\half\psbar_\mu\gamma_5\gamma^\lambda\psi_\lambda\ .\eqn\eeii$$

We now show that all the classical solutions found in Ref.\iii\ satisfy
the constraints and the equations of motion which are derived from the action
\ei .
In order to do this, we set all the fermionic fields to vanish, which
are solutions of all the constraints and the equations of motion which
are given by the variation of the fermionic fields $\psi_\mu$,
$\chi$ (or $\zeta$) and $\xi_i$.
Then by integrating the the auxiliary fields $S$, $F'$ (or $f'$) and
${G'}_i$,  we obtain the following classical action which has appeared in
the paper by Callan, Giddings, Harvey and Strominger\cref\iii
$$S=\pint\sqg \Bigl\lbrack
\ephi\Bigl(-R+4(\nabla \phi)^2+4\lambda^2\Bigr)
-\half\sum_i^N g^{\mu\nu}\partial_\mu a_i\partial_\nu a_i\Bigr\rbrack
\ . \eqn\eiii$$
This tells that all the classical solutions found in Ref.\iii , including the
solutions describing the formation of a black hole by collapsing matter, are
also solutions of this supersymmetric theory.

We note that we cannot construct a supersymmetric model of the dilaton gravity
when the cosmological constant $\lambda^2$ is negative.

\chapter{Quantum Effects}

In the original paper by CGHS, the quantum effects were expected to be
described by adding correction term, which only comes from the conformal
anomaly\cref\xxv to the classical action.
In the supersymmetric model proposed here, this corresponds to add
the following term\cref\xxvi\foot{
The local supersymmetric form of this action is given in Ref.\xxiv .}
$$S_{\rm anomaly}={\kappa \over 2\pi}\int d^2x\,
\Bigl\lbrack
-\half\partial_\mu\rho\partial^\mu\rho
-\half\psbar\gamma^\mu\partial_\mu\psi+\half e S^2\Bigr\rbrack
\ ,\eqn\eiv$$
when we choose the following superconformal gauge fixing condition,
$$g_{\mp\pm}=-\half\eprho\ , \ \ g_{\pm\pm}=0\ , \psi_\mu=\gamma_\mu\psi\ .
\eqn\ev$$
Here
$\kappa={8-N \over 4}$ is a constant, which should be determined by the
conformal anomaly.
We need more counterterms since the quantum action should have superconformal
symmetry when we choose the superconformal gauge \ev \pref\xxxi
By following de Alwis' paper\cref\xxxii we assume the kinetic term is given
by
$$\eqalign{S_{\rm kin}=&\pint d^2\theta \Bigl\lbrack
-4\e^{-2\hat\Phi}(1+h(\hat\Phi))\bar D \hat\rho D \hat\Phi\cr
&+2\e^{-2\hat\Phi}(1+\bar h(\hat\Phi))(\bar D \hat\Phi D\hat\rho
+\bar D \hat\rho D \hat\Phi)
+\kappa\bar D \hat\rho D \hat\rho\Bigr\rbrack
\ .}\eqn\evi$$
Here we have used superfield notations and $\hat\Phi$ and $\hat\rho$ are
superfields defined by
$$\eqalign{\hat\Phi \equiv&\phi+\bar\theta\chi+\half\bar\theta\theta F'\ , \cr
\hat\rho\equiv&\rho+\bar\theta\psi+\half\bar\theta\theta S\ .}\eqn\evii$$
$\theta$ and $\bar\theta$ are anti-commuting coordinates and $D$ and
$\bar D$ are covariant derivatives.
If we define new fields $\hat X$ and $\hat Y$ by
$$\eqalign{\hat X=&2\sqrt{2 \over |\kappa|}\int^{\hat\Phi}dt\e^{-2t}
\sqrt{(1+\bar h(t))^2+\kappa\e^{2t}(1+h(t))}\ ,\cr
\hat Y=&\sqrt{2|\kappa|}\Bigl(\hat\rho-{1 \over \kappa}
+{2 \over \kappa}\int^{\hat\Phi} dt
\e^{-2t}\bar h(t)\Bigr) \ .}\eqn\eviii$$
the kinetic term is rewritten by
$$S_{\rm kin}=\pint d^2\theta \lbrack
\mp\bar D \hat X  D \hat X \pm \bar D \hat Y D \hat Y\rbrack
\eqn\eix$$
Here upper$/$lower signs correspond to $\kappa >0\ /\ \kappa <0$,
respectively.
{}From now on, we will consider the case with lower signs in Eq.\eix\ since
another case can be treated in a similar way.

If we assume that there is any interaction term
with respect to $\hat X$ and $\hat Y$,
the energy momentum tensor $T$ has the following form,
$$\eqalign{T=&T_X+T_Y+T_\Sigma+T_{\rm ghost}\ ,\cr
T_X=&-\half(\partial X \partial X-\chi_X\partial \chi_X)\ ,\cr
T_Y=&\half(\partial Y \partial Y-\chi_Y\partial \chi_Y)
+\sqrt{|\kappa| \over 2}\partial^2 Y\ }\eqn\ex$$
Here we have written $\hat X$ and $\hat Y$ fields in the components
$\hat X=X+\bar\theta \chi_X+\half\bar\theta\theta F_X$ and
$\hat Y=Y+\bar\theta \chi_Y+\half\bar\theta\theta F_Y$.
$T_\Sigma$ and $T_{\rm ghost}$ are energy momentum tensors of matter
fields and ghost fields and they contribute to the central charge by
${3 \over 2}N$ and $-15$, respectively.
The contribution to the central charge by $T_X$ is ${3 \over 2}$ and that
by $T_Y$ is given by
$$c_Y={3 \over 2}(1+4\kappa)={3 \over 2}(-N+9)\ .\eqn\exi$$
Therefore the total central charge $c$ vanishes :
$c={3 \over 2}N-15+{3 \over 2}+c_Y=0$.
We now introduce interaction term $V$ so that the term does not
violate the superconformal symmetry.
This requires that $V$ should be given
by a vertex operator $V= :\e^{\alpha\hat X+\beta \hat Y}:$ whose conformal
dimension is $(\half,\half)$, \ie ,
$$\half\alpha^2-\half\beta(\beta+\sqrt{2|\kappa|})=\half\ .\eqn\exii$$
If we impose the condition that $T$ is proportional to $\e^{\rho-\phi}$ in
the weak coupling limit, we find $\alpha=\beta$, \ie ,
$$\alpha=\beta=\sqrt{1 \over 2|\kappa|}=\sqrt{2 \over |N-8|}\ .\eqn\exiii$$
Therefore we find the quantum theory is described by the following effective
action:
$$\eqalign{S_{\rm q}=&\pint d^2\theta \Bigl\lbrack
\bar D \hat X  D \hat X  - \bar D \hat Y D \hat Y
+2\tilde\lambda\e^{\sqrt{2 \over |N-8|}(\hat X+ \hat Y)}
+\sum_{i=1}^N\bar D \hat\Sigma D \hat\Sigma
\Bigr\rbrack\cr
=&\pint \Bigl\lbrack
-\Bigl(\partial_\mu X\partial^\mu X-
i\bar\chi_X\gamma^\mu\partial_\mu\chi_X
-F_X^2\Bigr)\cr
&+\Bigl(\partial_\mu Y\partial^\mu Y
-i\bar\chi_Y\gamma^\mu\partial_\mu\chi_Y-F_Y^2\Bigr)\cr
&+\tilde\lambda\sqrt{2 \over |N-8|}\e^{\sqrt{2 \over |N-8|}(X+Y)}
\Bigl\{F_X+F_Y\cr
&-\half\sqrt{2 \over |N-8|}(\bar\chi_X+\bar\chi_Y)(\chi_X+\chi_Y)\Bigr\}\cr
&+\sum_i^N\half\Bigl\{-\partial_\mu a_i\partial^\mu a_i
+i\xi_i\gamma^\mu\partial_\mu\xi_i+{G}_i^2\Bigr\}\Bigr\rbrack
}\eqn\exiv$$
Here $\hat\Sigma_i$'s are matter superfields:
$\hat\Sigma_i=a_i+\bar\theta \xi_i+\half\bar\theta\theta G_i$.
By integrating auxiliary fields $F_X$, $F_Y$ and $G_i$,
we obtain,
$$\eqalign{{S'}_{\rm q}
=&\pint \Bigl\lbrack
-\Bigl(\partial_\mu X\partial^\mu X
-i\bar\chi_X\gamma^\mu\partial_\mu\chi_X\Bigr)\cr
&+\Bigl(\partial_\mu Y\partial^\mu Y
-i\bar\chi_Y\gamma^\mu\partial_\mu\chi_Y\Bigr)\cr
&-{\tilde\lambda \over |N-8|}\e^{\sqrt{2 \over |N-8|}(X+Y)}
\sqrt{2 \over |N-8|}(\bar\chi_X+\bar\chi_Y)(\chi_X+\chi_Y)\cr
&+\sum_i^N\{-\partial_\mu a_i\partial^\mu a_i
+i\xi_i\gamma^\mu\partial_\mu\xi_i\}\Bigr\rbrack
}\eqn\exv$$
Note that terms like $\e^{2\sqrt{2 \over |N-8|}(X+Y)}$ do not appear.
When we consider more general interaction term
$V= :\e^{\alpha\hat X+\beta \hat Y}:$, which satisfy Equation \exii\
and $\alpha\neq\beta$, the term like
$V= :\e^{2(\alpha\hat X+\beta \hat Y)}:$ does not appear as long as $V$
is exactly marginal, \ie , $V(x)V(y)\sim O((x-y)^\delta)$, $\delta>0$, which
tel
ls
$V(x)^2=0$.

The equations of motion for $X$ and $Y$ are given by
$$\eqalign{0=&\partial_\mu \partial^\mu X
-{\tilde \lambda \over 2}\Bigl({2 \over |N-8|}\Bigr)^{{3 \over 2}}
\e^{\sqrt{2 \over |N-8|}(X+Y)}(\bar\chi_X+\bar\chi_Y)(\chi_X+\chi_Y)\cr
0=&-\partial_\mu \partial^\mu Y
-{\tilde\lambda \over 2}\Bigl({2 \over |N-8|}\Bigr)^{{3 \over 2}}
\e^{\sqrt{2 \over |N-8|}(X+Y)}(\bar\chi_X+\bar\chi_Y)(\chi_X+\chi_Y)
}\eqn\exvi$$
If we consider solutions where all the fermion fields vanish,
$X$ and $Y$ are given by the sums of holomorphic and anti-holomorphic
functions. This tells that there is not any solution corresponding to
black hole solution in the classical theory.
Supersymmetry forbids the existence of black hole
in the quantum theory.

If we bosonize the fermion fields $\chi_X$ and $\chi_Y$:\foot{
Since the fermion fields $\chi_X$ and $\chi_Y$ have opposite
signatures, these fermion fields can be bosonized by a negative norm
boson\pref\xxxx}
$\chi_X\pm\chi_Y\sim\pm:\e^{\pm \vartheta}$:, we obtain the following
equations of motion,
$$\eqalign{0=&\partial_\mu \partial^\mu X
-{\tilde\lambda \over 2}\Bigl({2 \over |N-8|}\Bigr)^{{3 \over 2}}
\e^{\sqrt{2 \over |N-8|}(X+Y)+\vartheta}\cr
0=&-\partial_\mu \partial^\mu Y
-{\tilde \lambda \over 2}\Bigl({2 \over |N-8|}\Bigr)^{{3 \over 2}}
\e^{\sqrt{2 \over |N-8|}(X+Y)+\vartheta}\cr
0=&\partial_\mu \partial^\mu \vartheta
-{\tilde \lambda \over 2}\Bigl({2 \over |N-8|}\Bigr)^{{3 \over 2}}
\e^{\sqrt{2 \over |N-8|}(X+Y)+\vartheta}}\eqn\exvii$$
We can set $X=Y$ by using the residual symmetry of the reparametrization
symmetry or by a coordinate choice.
Then we find that $\vartheta$ satisfies the Liouville equation,
$$0=\partial_\mu \partial^\mu \vartheta
-{\tilde \lambda\over 2}\Bigl({2 \over |N-8|}\Bigr)^{{3 \over 2}}
\e^\vartheta\ .
\eqn\exviii$$
The equations \exvii\ tell that $X$ and $Y$ are given in terms of $\vartheta$,
$$X=-Y=\vartheta+f^+(x^+)+f^-(x^-)\ .\eqn\exix$$
Here $f^\pm$ are arbitrary functions.
Note that there are static solutions:
$$f^\pm=0\ , \ \
\e^{-\vartheta}=-{16 \over A}{Cn^2(x^+x^-)^{n-1} \over \{1-C(x^+x^-)^n\}^2}\ .
\eqn\exx$$
Here $A=\tilde \lambda\Bigl({2 \over |N-8|}\Bigr)^{{3 \over 2}}$, $C$ is an
arbitrary constant and $n$ is an integer.

\chapter{Summary and Discussion}

We have analyzed a supergravity theory coupled to a dilaton and superconformal
matters in two dimensions.
This theory is classically soluble and we have found
all the solutions appeared in Callan, Giddings, Harvey and Strominger's
dilatonic gravity also satisfy the constraints and the equations of motion in
this supersymmetric theory.
When we quantize this theory following the procedure of Distler, Hlousek
and Kawai, the cosmological term is renormalized to vanish in the quantum
action.
As a result, any solution corresponding to classical black hole
does not appear in the quantum
theory, which should be compared with the non-supersymmetric case.

It should be amazing that supersymmetry forbids the existence of quantum
black hole although classical black hole is allowed to exist.
One of the motivations of the present work was to build a theory of dilaton
gravity where the Bondi mass of black hole is bounded from below and the
theory has a ground state. The above motivation becomes, however, irrelevant
since there is no quantum black hole in our model.

However, from the alternative viewpoint, our model gives us an interesting
conjecture.
Let us assume that the quantum black hole disappears even in four dimensions
although it is difficult to take account of quantum effects owing to the
non-renormalizability of gravity. In four dimensions, the Schwarzschild
radius of black holes and the Compton wave length of elementary particles
become comparable at the Planck length scale. This might suggest that
we should include the black hole--like states in the Hilbert space. From
the observation done in the present article, however, we might conjecture
that, if supersymmetry is realized at the Planck scale, such black hole--like
states need not to be included as quantum states in quantum gravity where
only quantum states expressing smooth space-time structure are admitted.


We wish to thank K. Higashijima, M. Kato, K. Odaka and A. Sugamoto for valuable
discussion. The work of I.O. is supported by the Japan Society for the
Promotion

of Science.

\refout

\bye